\documentclass[aps]{revtex4}
\usepackage{graphicx}
\usepackage{amsmath}
\usepackage{amsfonts}
\usepackage{amssymb}

\newcommand{\beq}{\begin{equation}}
\newcommand{\eeq}{\end{equation}}
\newcommand{\beqa}{\begin{eqnarray}}
\newcommand{\eeqa}{\end{eqnarray}}
\newcommand{\eps}{\epsilon}
\newcommand{\rr}{{\bf r}}

\newcommand{\p}{{\bf p}}

\begin{document}

\title{Exact Tkachenko modes and their damping in the vortex lattice regime of rapidly  rotating bosons}
\author{ S. I. Matveenko$^1$  and  G. V. Shlyapnikov$^{2,3,4}$}
\affiliation{\mbox{$^1$L.D. Landau Institute for Theoretical Physics, Kosygina Str. 2, 119334, Moscow, Russia}\\
\mbox{$^2$ Laboratoire de Physique Th\'eorique et Mod\'eles Statistiques, Universit\'e Paris Sud, CNRS, 91405~Orsay, France} \\
\mbox{$^3$ Van der Waals-Zeeman Institute, University of Amsterdam, Valckenierstraat 65/67, 1018 XE~ Amsterdam, The Netherlands}\\
\mbox{$^4$ Kavli Institute for Theoretical Physics, University of California, Santa Barbara, California 93106-4030, USA}
}
\date{\today}

\begin{abstract}
We have found an exact analytical solution of the Bogoliubov-de Gennes equations for the
  Tkachenko modes of the vortex lattice in the lowest Landau level (LLL) in the
thermodynamic limit at any momenta and calculated their damping rates. At finite
temperatures both Beliaev and Landau damping leads to momentum independent damping rates in the
low-energy limit, which shows that at sufficiently low energies Tkachenko modes become
strongly damped. We then found that the mean square fluctuations of the density grow
logarithmically at large distances, which indicates that the state is ordered in the vortex
 lattice only on a finite (although exponentially large) distance scale and introduces a
low-momentum cut-off. Using this circumstance we showed that at finite temperatures the
 one-body density matrix undergoes an exponential decay at large distances.
\end{abstract}
\pacs{03.75.Lm, 05.30.Jp, 73.43.Nq}

\maketitle
\section{Introduction}

The physics of rapidly rotating Bose-condensed gases is governed by a collective behavior of
nucleated vortices (see \cite{Zwerger,coop,revfet} for review). When the rotation
frequency $\Omega$ along the axis perpendicular to the plane of rotation $(x , y)$ is close
to the trapping frequency $\omega$ in the $x$ and $y$ directions, the harmonically
trapped Bose gas becomes essentially two-dimensional (2D) and can be  described  as a
 system of interacting bosons in the lowest Landau level (LLL). If the number of vortices is
much smaller than the number of particles, the system is in the so-called "mean-field"
 quantum Hall regime \cite{ho}, where vortices arrange themselves in a lattice with an
intervortex spacing of the order of the "magnetic length" $l = \hbar/\sqrt{m \Omega}$.
Under an increase  in $\Omega$ the vortex lattice should melt and strongly correlated
Quantum Hall states should emerge.

The vortex lattice in the LLL or close to this regime has been obtained experimentally
  \cite{exp1,exp2,exp3} and studied theoretically assuming the presence of a macroscopic wave
function $\Psi(\bf{r})$ (see \cite{revfet} for review and refs. in \cite{us}). Presently, a
new generation of experiments with rapidly rotating quantum gases is being set up. It
is based on the use of  artificial gauge potentials, which are  obtained by means of
 combinations of laser fields and mimic the rotation of the system (see \cite{Zwerger} for
review). They are supposed to provide the  "rotation"  with frequency that is extremely close
 to $\omega$, thus allowing for the  observation of the lattice melting and for the
emergence of strongly correlated states. This puts forward the question of correlation
 properties in the   vortex lattice regime and revives the interest to excitation modes in
this regime.

The theory of elastic oscillations of a vortex lattice in incompressible superfluids has been
 constructed by Tkachenko \cite{tk}, and a linear spectrum of the wave dispersion has
been obtained. The effect of a finite compressibility has been discussed in
Refs. \cite{vol-doz,son,baym1}, and it has been shown that the compressibility changes dramatically the
dispersion relation in the low-momentum limit. The dispersion becomes quadratic due to
 hybridization with sound waves. The first experimental observation of Tkachenko modes in
harmonically trapped Bose-condensed gases has been reported in Ref.\cite{exp1}. Theoretical
 analysis of these modes has been done in the mean-field approach for both the geometry
of an infinite plane ($\Omega=\omega$) and in the presence of remaining
 trapping \cite{anglin,baym2,baym3,sonin1,sonin2,mizushima,baksmaty,cozzini}. The calculations provided an
explanation of the mode frequencies observed in the experiment \cite{exp1}. Theoretical
 studies based on the microscopic effective action showed the absence of long-range order at
$T=0$ in the thermodynamic limit (infinite plane geometry) \cite{sinova}, and  the hydrodynamic
 approach revealed an algebraic decay of the one-body density matrix at large
distances \cite{baym3}.

In this paper we find an exact analytical solution of the Bogoliubov-de Gennes equations for the
  Tkachenko modes of the LLL vortex lattice in the thermodynamic limit
at any momenta and calculate their damping rates. Importantly, at finite temperatures both
 Beliaev and Landau damping mechanisms lead to momentum independent damping rates in the
low-energy limit. This means that for sufficiently low energies the Tkachenko modes become
 strongly damped, which is consistent with the experimental results \cite{exp1}. Using the
obtained results for the excitation wavefunctions we then calculate the mean square fluctuations
 of the density and the one-body denity matrix. The density fluctuations grow
logarithmically at large distances, which indicates that the state is ordered in the vortex lattice only on a finite (although exponentially large) distance scale and introduces a
low-momentum cut-off. Using this circumstance we show that at finite temperatures the one-body density matrix undergoes an exponential decay at large distances.

\section{The  ground state  and equations for the excitations in the LLL approximation. }

We consider a zero-temperature two-dimensional  (2D) system of  bosonic  atoms in  a harmonic
 trapping potential $V(r)=m\omega^2r^2/2$,  rotating  with frequency $\Omega$ around
the axis perpendicular to the $(x, y)$ plane. In the rotating frame the Hamiltonian has the form:
\beq
H = \int d^2 {\bf{r}} \, \left[\hat{\psi}^{\dagger}\frac{\hat{\p}^2}{2 m} \hat{\psi} + \frac{g}{2} \hat{\psi}^{\dagger}\hat{\psi}^{\dagger} \hat{\psi}\hat{\psi} +V(\rr)
\hat{\psi}^{\dagger}\hat{\psi}
 - \Omega \hat{\psi}^{\dagger}\hat{L}_z \hat{\psi}\right],
 \eeq
where $\hat{\psi}({\bf{r}})$ is the  field  operator,  $\hat{\p}$ is the momentum operator, $m$
 is the atom mass, $\hat{L}_z$ is the operator of the orbital angular momentum, and
$g$ is the coupling constant for short-range atom-atom interaction. The non-linear
 Schroedinger equation for $\hat\psi({\bf r},t)$ reads:
\begin{equation}         \label{Schrt}
i\hbar\frac{\partial\hat\psi}{\partial t}=\frac{\hat{\p}^2}{2 m} \hat{\psi} + g\hat{\psi}^{\dagger} \hat{\psi}\hat{\psi} +V(\rr)\hat{\psi}
 - \Omega\hat{L}_z \hat{\psi}.
\end{equation}
It is commonly assumed   that in the mean-field Quantum Hall regime  all particles are in the
 same macroscopic quantum state described by the wavefunction $\Psi$.
\cite{Zwerger,coop,revfet,ho}. In the ground state this wavefunction has the form
 $\Psi({\bf r},t)=\Psi_0({\bf r})\exp(-i\mu t)$, where $\mu$ is the chemical potential. The
wavefunction $\Psi_0({\bf r})$ is then governed by the Gross-Pitaevskii equation:
\beq
\frac{\hat{\p}^2}{2 m} \Psi_0 + g |\Psi_0|^2 \Psi_0  +V(\rr) \Psi_0  - \Omega \hat{L}_z \Psi_0 = \mu \Psi_0,
\label{gp}
\eeq
where it is normalized to the total number of particles $N$. For $\Omega$ close to $\omega$,
 the ground state of the system corresponds to the LLL and, hence, the macroscopic wave
function $\Psi_0(\bf{r})$ takes the form:
\beq
\Psi_0({\bf{r}}) = \sqrt{n} f_0(z) \mbox{e}^{- |z|^2 /2},
\eeq
where $n=N/S$ is the mean density, with $S$ being the surface area,
 $z, \bar{z} = (x \pm i y)/l$, and the function $f_0(z)$ is analytical in the $(x, y)$ plane, with $f_0(z) \exp
(-|z|^2/2)$ normalized to unity. Equation (\ref{gp})  is then solved numerically by expressing
 $\Psi_0$ as a superposition of the LLL single-particle eigenfunctions
 $z^n \exp(-|z|^2/2) /\sqrt{\pi n!}$. Alternatively, one solves a projected Gross-Pitaevskii
  equation obtained by  acting on Eq. (\ref{gp}) with the LLL projection operator $\hat
P$. This  operator acts on  an arbitrary function $F(z, \bar{z})$ as
\beq
\hat{P} F(z, \bar{z}) = \frac{1}{\pi} \int dw d \bar{w} \exp [-|w|^2 + z \bar{w}] F(w, \bar{w}),
\eeq
and transformes it to the  function in the LLL.

For $\Omega = \omega$ we have the geometry of an infinite plane. The LLL is infinitely degenerate and the projected Gross-Pitaevskii equation reads
\beq
\frac{N g}{\pi}  \int dw d\bar{w} \mbox{e}^{-2 w \bar{w} + z \bar{w}} |f_0(w)|^2 f_0(w)
= \tilde\mu  f_0(z),
\label{P}
\eeq
where $\tilde\mu=\mu-\hbar\Omega$.

The ground state solution of Eq. (\ref{P}) is a triangular  vortex lattice and the function
 $f_0(z)$ is expressed through the Jacobi Theta-function $\vartheta_1$ \cite{us}:
\beq
f_0 (z) =  (2v)^{1/4}\vartheta_1 (\sqrt{\pi v} z , q )\, \mbox{e}^{z^2 /2},
\eeq
with $q = \exp(i \pi \tau)$, $\tau = u + i v$, $v = \sqrt{3}/2$,  $u = - 1/2$. The chemical
 potential is then equal to $\tilde\mu=\alpha ng$, with $\alpha=\sqrt{v}\sum_{m,p}
(-1)^{mp}\exp\{-\pi v(m^2+p^2)\}=1.1596$.

An alternative mean-field approach does not apriori assume the presence of long-range order and
 is based on the density-phase formalism. It is commonly  employed for 2D
Bose-condensed gases  at finite temperatures  and for weakly interacting 1D bosons, where the
 long-range order is destroyed by long-wave  fluctuations of the  phase. The key
condition of this approach is related to small fluctuations of the density. One writes the field operator in the form
\beq
\hat{\psi} = \exp i \hat\Phi\, \sqrt{\hat{n}};\,\,\,\,\,\,\,\hat\psi^{\dagger}= \sqrt{\hat{n}}\exp{(-i\hat\Phi)}
\label{dfr}
\eeq
with $\hat{n}$ and $\hat{\Phi}$ being the density and phase operators satisfying the the commutation relation
 $[\hat{n}(\rr), \hat{\phi}(\rr^{\prime})] = i\delta (\rr -\rr^{\prime})$. Representing these operators as
 $n = n_0({\bf{r}}) + \delta \hat n$ and $\Phi =\Phi_0({\bf{r}}) + \delta \hat \Phi$, one  writes
  Eq. (\ref{Schrt}) in terms of the density and phase  keeping only zero and first
order terms in small fluctuations $\delta\hat n$ and $\nabla\delta \hat\Phi$. To zero order
 we then have Eq.~(\ref{Schrt}) for $\Psi_0({\bf r}) = \sqrt{n_0 ({\bf r})} \exp
(i \Phi_0 ({\bf r}))$
and, hence, the projected equation (5) for the function $f_0(z)$  introduced by Eq.~(4). The
 solutions of two equations that are linear in $\delta \hat{n}$ and $\nabla\delta
\hat\Phi$ are obtained by representing these quantities in terms of elementary excitations
  characterized by the wavenumber ${\bf k}=\{k_x,k_y\}$:
\beq
\delta\hat n =\sqrt{n_0}\mbox{e}^{-|z|^2/2}\sum_{{\bf k}} [u_{{\bf k}} \exp[-i\Phi_0]
 -\tilde{v}^*_{{\bf k}} \exp[i\Phi_0]] \exp[-i\eps_{{\bf k}} t]\hat{a}_{{\bf k}} + h.c.
\label{dn}
\eeq
\beq
\delta\hat\Phi = \frac{-i \mbox{e}^{-|z|^2/2}}{2\sqrt{n_0}}
\sum_{{\bf k}}[u_{{\bf k}}\exp[-i\Phi_0] +\tilde{v}^*_{{\bf k}} \exp[i\Phi_0]]
 \exp[-i\eps_{{\bf k}} t]\hat{a}_{{\bf k}}
+
h.c. ,
\label{df}
\eeq
where  $\hat{a}_{{\bf k}}$, $\hat{a}_{{\bf k}}^{\dagger}$ are operators of anihilation/creation of the excitations, and the
functions $u_{{\bf k}},\tilde v_{{\bf k}}$ satisfy the Bogoliubov-de
Gennes equations. In the LLL approximation they are analytical functions of $z$ and follow from
 the projected Bogoliubov-de Gennes equations:
\beqa
2 g \hat{P} (|\Psi_0|^2 u_{{\bf k}}) - g \hat{P}(\Psi_0^2 \tilde{v}^*_{{\bf k}}) &=& (\tilde\mu + \eps_{{\bf k}} ) u_{{\bf k}}  \nonumber \\
2 g \hat{P} ( |\Psi_0|^2 \tilde{v}_{{\bf k}}) - g \hat{P} (\Psi_0^2 {u}^*_{{\bf k}}) &=& (\tilde\mu - \eps_{{\bf k}} ) \tilde{v}_{{\bf k}}.
\label{BE}
\eeqa

\section{Exact solution of the projected Bogoliubov-de Gennes equations}

In this section we will use coordinates expressed in units of $l$, wavevectors in units of $l^{-1}$,
 and energies in units of $ng$.
The exact solution of  Eqs.~(\ref{BE}) for the functions $u_{{\bf k}},\,\tilde v_{{\bf k}}$  reads:
\beqa
u_{{\bf k}} = \frac{c_{1{\bf k}}}{\sqrt{S}} f_0 \left(z + \frac{i k_{+}}{2}\right)
\mbox{e}^{i k_- z/2} \mbox{e}^{-k^2/4} = c_{1{\bf k}} P (f_0 \mbox{e}^{i \bf{k r}})   \\
\tilde{v}_{{\bf k}} = \frac{c_{2{\bf k}}}{\sqrt{S}} f_0 \left(z - \frac{i k_{+}}{2}\right)
\mbox{e}^{-i k_- z/2}\mbox{e}^{-k^2/4} = c_{2{\bf k}} P (f_0 \mbox{e}^{-i \bf{kr}}),
\eeqa
where $k_{\pm} = k_x  \pm  i k_y$, and the coefficients $c_{1{\bf k}},\,c_{2{\bf k}}$ follow from
 the normalization condition $\int dx \, dy (|u_{{\bf k}}|^2 -|v_{{\bf k}}
^2)\mbox{e}^{-|z|^2} = 1$. They are given by
\beq
c_{1{\bf k}} =  \left[\frac{\tilde{K}({\bf k}) + \eps_{{\bf k}}}{2 \eps_{{\bf k}}}\right]^{1/2}\mbox{e}^{k^2/8},
\label{c1k}
\eeq
\beq
c_{2{\bf k}} = \left[\frac{\tilde{K}({\bf k}) - \eps_{{\bf k}}}{2 \eps_{{\bf k}}}\right]^{1/2} \frac{|K_2({\bf k})|}{ K_2({\bf k})} \mbox{e}^{k^2/8},
\label{c2k}
\eeq
with
\beq
\tilde{K}({\bf k}) = 2 K_1({\bf k}) - K_1(0),
\eeq
\beq
K_1({\bf{k}}) = \sqrt{v}\sum_{n,m = -\infty}^{\infty}(-1)^{n m} \mbox{e}^{- \pi v (n^2 + m^2)}
\mbox{e}^{-\sqrt{\pi v} k_x n + i \sqrt{\pi v} k_y m }\mbox{e}^{-k_x^2/4},
\label{k1}
\eeq
\beq
K_2({\bf{k}}) = \sqrt{v}\sum_{n,m = -\infty}^{\infty}(-1)^{n m} \mbox{e}^{- \pi v (n^2 + m^2)}
\mbox{e}^{-\sqrt{\pi v} (k_x - i k_y)( n + m) }\mbox{e}^{-k_x^2/2 + i k_x k_y /2},
\label{k2}
\eeq
and
\beq
K_1(0) = K_2(0)= \tilde K(0)=\alpha\simeq 1.1596.
\eeq
The sums in Eqs.~(\ref{k1}) and (\ref{k2}) can be expressed in terms of Jacobi Theta-functions:
\beq
K_1({\bf{k}}) =\frac{e^{-\frac{{k_y}^2}{4}} \left(\vartheta _3\left(\frac{{k_x} \sqrt{\pi }}{2 \sqrt{v}},e^{-\frac{\pi }{v}}\right) \vartheta _3\left(-\frac{i {k_y} \sqrt{\pi }}{4
\sqrt{v}},e^{-\frac{\pi }{4 v}}\right)+\vartheta _2\left(\frac{{k_x} \sqrt{\pi }}{2 \sqrt{v}},e^{-\frac{\pi }{v}}\right) \vartheta _4\left(-\frac{i {k_y} \sqrt{\pi }}{4
\sqrt{v}},e^{-\frac{\pi }{4 v}}\right)\right)}{2 v}
\eeq
\beq
\begin{split}
K_2({\bf{k}}) = \frac{1}{2 v}e^{-\frac{1}{2} {k_y} ({k_y}+i {k_x})} \bigg[\vartheta _3\left(\frac{({k_x}-i {k_y}) \sqrt{\pi }}{2 \sqrt{v}},e^{-\frac{\pi }{v}}\right) \vartheta
_3\left(\frac{({k_x}-i {k_y}) \sqrt{\pi }}{4 \sqrt{v}},e^{-\frac{\pi }{4 v}}\right) \nonumber \\
+\vartheta _2\left(\frac{({k_x}-i {k_y}) \sqrt{\pi }}{2 \sqrt{v}},e^{-\frac{\pi }{v}}\right) \vartheta _4\left(\frac{({k_x}-i {k_y}) \sqrt{\pi }}{4 \sqrt{v}},e^{-\frac{\pi }{4
v}}\right)\bigg]
\end{split}
\eeq

The spectrum of excitations has the form (see Fig.~1 and Fig.~2):
\beq                 \label{dr}
\eps_{{\bf k}}^2 = | 2 K_1({\bf{k}}) - K_0|^2 - |K_2 ({\bf{k}})|^2,
\eeq
\begin{figure}[htb]
 \includegraphics[width=3.0in]{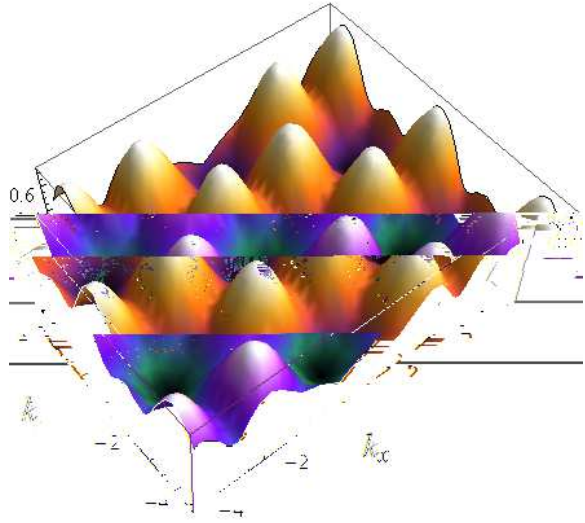}
 \caption{Excitation energy $\eps(k_x, k_y)$ in units of $ng$.}
\label{fig1}
 \end{figure}
 \begin{figure}[htb]
 \includegraphics[width=3.0in]{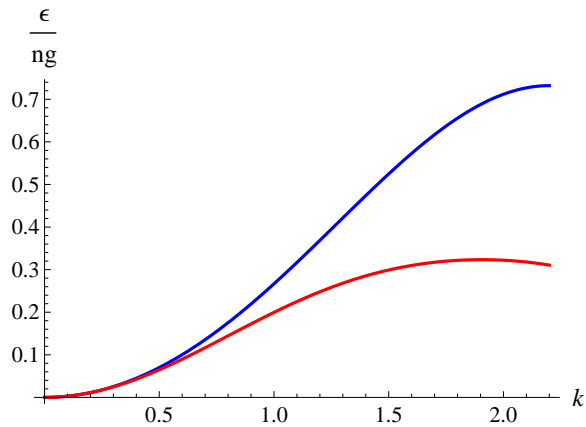}
 \caption{Anisotropy of the dispersion relation (\ref{dr}). The upper curve is $\eps(k, 0)$,  the
  lower curve is $\eps(k \cos\frac{\pi}{6}, k\sin \frac{\pi}{6})$, and the
directions $x$ and $y$ are chosen in the same way as in Fig.~1.    }
\label{fig2}
 \end{figure}
It is consistent with the spectrum obtained in the hydrodynamic approach
 \cite{baym2,baym3,sonin2}, but we clearly identify the anisotropy: for a given $k$ the excitation energy is
maximal in the $x$ direction, and minimal at an angle of 30 degrees from the $x$ 
axis. (see Fig.~2). In the limit of small $k$, using the expansion of the functions $K_1({\bf k})$
and $K_2({\bf k})$:
\beq
K_1 = \alpha\left[1 - \frac{k^2}{8} + \frac{(\eta+1) k^4}{64}\right]; \quad K_2 = \alpha\left(1 - \frac{k^2}{4} + \frac{k^4}{32}\right),
\eeq
where
$$\eta =-8\,\frac{\sum_{m,p}(-1)^{mp}m^2p^2\exp[-\pi v(m^2+p^2)]}{\sum_{m,p}(-1)^{mp}\exp[-\pi v(m^2+p^2)]}\simeq 0.8219,$$
we find a symmetric spectrum. Restoring the dimensions it reads:
\beq
\eps=\frac{\alpha\sqrt{\eta}}{4}\,ng  (k l)^2\simeq 0.2628\, ng\, (kl)^2,
\label{lowkspectrum}
\eeq
which exactly coincides with the hydrodynamic result of Ref.~\cite{sonin2}. The obtained
 excitations are commonly termed as Tkachenko modes, although they have a quite different
dispersion relation compared to elastic oscillations of the vortex lattice in incompressible superfluids obtained by Tkachenko \cite{tk}.

For atoms (tightly) confined to the quasi2D geometry with frequency $\omega_0$, the coupling
 constant for the interatomic interaction is $g=2\sqrt{2\pi}\hbar^2a/ml_0$, where
$a$ is the 3D scattering length, and $l_0=\sqrt{\hbar/m\omega_0}$. In the case of $^{87}$Rb
 rotating in the $(x,y)$ plane with frequency $\Omega\simeq 100$ Hz and confined in the
perpendicular direction with  $\omega_0\simeq 300$ Hz, one has $ng/\hbar\Omega\simeq 0.1$ at
 the 2D density $n\simeq 3\times 10^8$ cm$^{-2}$, which justifies the LLL
approximation. Then, low-energy excitations have frequencies below 1 Hz. Note that for this
 example the quantity $\nu=\pi nl^2$ representing the ratio of the number of particles
to the number of vortices, i.e. the filling factor, is large and we are well in the mean-field regime.

\section{Damping rates of the excitations}

Let us now calculate the damping of these excitations, which is caused by the interaction term of
 the Hamiltonian (1), containing a product of four field operators. For finding the
damping rate it is sufficient to use a linearized form of the field operator, i.e. put $\hat\psi=\Psi_0(1+\delta\hat n/2n_0+i\delta\hat\Phi)$. Using equations (4), (\ref{dn}) and
(\ref{df}) we then have:
\begin{equation}         \label{psilin}
\hat\psi=\left(\sqrt{n}f_0(z)+\sum_{{\bf k}}[u_{{\bf k}}\hat a_{{\bf k}}\exp(-i\epsilon_{{\bf k}}t)-\tilde v_{{\bf k}}\hat a^{\dagger}_{{\bf k}}\exp(i\epsilon_{{\bf k}}t)\right)
\exp(-i\mu t-|z|^2/2).
\end{equation}

We first consider the Beliaev damping mechanism \cite{LL9} in which a given excitation with wavevector ${\bf p}$ decays into two excitations
with lower energies and wavevectors (${\bf k}$ and ${\bf q}$). The part of the interaction
 Hamiltonian that causes the Beliaev damping contains three operators of the excitations
and has the form:
\beq
\hat{V} = g\sqrt{n}\sum_{{\bf k},{\bf q}}\int dx\, dy \left[f_0 u_{{\bf p}}u_{{\bf k}}^* u_{{\bf q}}^*  + 2 f_0 \tilde{v}_{{\bf p}}^* u_{{\bf q}}^* \tilde{v}_{{\bf k}} -f_0^*
\tilde{v}_{{\bf p}}^*
\tilde{v}_{{\bf k}} \tilde{v}_{{\bf q}} -2 f_0^* u_{{\bf p}}u_{{\bf k}}^* \tilde{v}_{{\bf q}}\right]
\mbox{e}^{-2 |z|^2} \hat{a}_{{\bf q}}^{\dagger} \hat{a}_{{\bf k}}^{\dagger} \hat{a}_{{\bf p}}\,+h.c.
\eeq
At a finite temperature $T$ we have to take into account thermal occupation of the states with
 momenta ${\bf k}$ and ${\bf q}$ and the presence of the reversed process in which
excitations with momenta ${\bf k}$ and ${\bf q}$ recombine into the excitation with momentum ${\bf p}$.
Using the Fermi golden rule the damping rate is given by
\beq
\Gamma_{{\bf p}}=\frac{2 \pi}{\hbar}\sum_{{\bf k},{\bf q}} |\langle {\bf k},{\bf q}|\hat V|{\bf p}\rangle|^2(1+N_{{\bf k}}+N_{{\bf q}}) \delta(\epsilon_{{\bf p}}-\epsilon_{{\bf
k}} -\epsilon_{{\bf q}}),
\label{Gamma}
\eeq
with $N_{{\bf k},{\bf q}}=[\exp(\epsilon_{{\bf k},{\bf q}}/T)-1]^{-1}$ being equilibrium occupation numbers for the excitations.
The excitation energy thus acquires the imaginary part and becomes $\epsilon_{{\bf p}}-i\hbar\Gamma_{{\bf p}}/2$.

For performing the calculations it is convenient to use the functions $u_{{\bf k}},\,v_{{\bf k}}$ in the form of Bloch waves:
\begin{eqnarray}
&&u_{{\bf k}}({\bf r})\exp(-|z|^2/2)=c_{1{\bf k}}\frac{\Psi_0(z+ik_+l/2)}{\sqrt{N}}\exp(-k^2l^2/8)\exp(i{\bf k}{\bf r}/2),  \label{uB} \\
&&v_{{\bf k}}({\bf r})\exp(-|z|^2/2)=c_{2{\bf k}}\frac{\Psi_0(z-ik_+l/2)}{\sqrt{N}}\exp(-k^2l^2/8)\exp(-i{\bf k}{\bf r}/2).  \label{vB}
\end{eqnarray}
In the low-energy limit where $pl\ll 1$, the transition matrix element is equal to
\beqa
\langle {\bf k},{\bf q}|\hat V|{\bf p}\rangle=\frac{\alpha^{5/2}}{S}\sqrt{\frac{N ng}{8\epsilon_p\epsilon_k\epsilon_q}}\,\left\{\frac{\epsilon_k}{\tilde
K(k)}+\frac{\epsilon_q}{\tilde K(q)}-\frac{\epsilon_p}{\tilde K(p)}\right\}\,\delta_{{\bf p},{\bf k}+{\bf q}},
\label{Vinterm}
\eeqa
where $\delta_{{\bf p},{\bf k}+{\bf q}}$ is the Kronecker symbol. After a straightforward algebra Eq.~(\ref{Vinterm}) is reduced to
\begin{equation}       \label{Vfin}
\langle {\bf k},{\bf q}|\hat V|{\bf p}\rangle=\frac{\alpha}{4\sqrt{2}\,\eta^{1/4}}\sqrt{\frac{ng^2}{S}}\,\frac{(k^4+q^4-p^4)l}{kqp}\,\delta_{{\bf p},{\bf k}+{\bf q}}.
\end{equation}
Equation (\ref{Gamma}) then yields:
$$\Gamma_p=\frac{\alpha^2}{8\sqrt{\eta}}\frac{ng^2l^2}{\hbar}\int_0^pkdk\int_0^{2\pi}\frac{d\phi}{2\pi}\,\frac{k^2(p^2-k^2)}{p^2}\coth\left(\frac{\epsilon_k}
{T}\right)\delta\left[\frac{\alpha\sqrt{\eta}}{4}ngl^2(p^2-k^2-(p^2+k^2-2kp\cos\phi)\right],$$
and we obtain:
\begin{equation}         \label{Gammafingen}
\Gamma_p=\frac{\alpha}{8\pi\eta}\,\frac{g}{\hbar}\,p^2\int_0^1 dx x^2\sqrt{1-x^2}\,\coth\left(\frac{\epsilon_p}{2T}\,x^2\right).
\end{equation}

At $T=0$ we immediately find:
\begin{equation}     \label{Gammafin0}
\Gamma_{p0}=\frac{\alpha}{128\eta}\,\frac{g}{\hbar}\,p^2\simeq 0.011\frac{g}{\hbar}\,p^2.
\end{equation}
For the ratio of the damping rate to the excitation energy we then obtain:
\begin{equation}        \label{RGE}
\frac{\hbar\Gamma_{p0}}{\epsilon_p}=\simeq \frac{0.13}{\nu}.
\end{equation}

In the mean-field regime we should have $\nu=\pi nl^2\gg 1$, since this quantity (filling factor)
 represents the ratio of the number of particles to the number of vortices.
Therefore, we  have $\hbar\Gamma_{p0}\ll\epsilon_p$ at any $p$. Thus, at $T=0$ Tkachenko
 modes are good elementary excitations in the entire mean-field Quantum Hall regime.

The situation changes drastically at finite temperatures. Equation (\ref{Gammafingen}) yields
 $\Gamma_{p0}+\Gamma_{pT}$, and for $\epsilon_p\ll T$ the temperature-dependent part of
the damping rate is independent of $p$ and proves to be
\begin{equation}        \label{GammafinT}
\Gamma_{pT}=\frac{\pi}{4\eta^{3/2}}\,\frac{T}{\hbar}\,\frac{1}{\nu}\simeq 1.05\frac{T}{\hbar}\,\frac{1}{\nu};\,\,\,\,\,\,\epsilon_p\ll T.
\end{equation}
For excitation energies $\epsilon_p\gg T/\nu$ we have $\hbar\Gamma_{pT}\ll\epsilon_p$, i.e. the
 damping rate is small and, hence, Tkachenko modes are good elementary
excitations. Moreover, for $\epsilon_p> T$ the damping rate starts to decrease with increasing
 $p$. In particular, for $\epsilon_p\gg T$ equation (\ref{Gammafingen}) gives:
\begin{equation}        \label{GammafinTbig}
\Gamma_{pT}=\frac{\sqrt{\pi}}{4\eta^{3/2}}\,\zeta(3/2)\frac{T}{\hbar\nu}\left(\frac{T}{\epsilon_p}\right);\,\,\,\,\,\epsilon_p\gg T,
\end{equation}
where $\zeta(3/2)$ is the Riemann zeta-function, and the imaginary part of the excitation
 energy is negligible compared to the real part $\epsilon_p$. However, excitations with
energies
\begin{equation}         \label{epsilonc}
\epsilon_p\lesssim\epsilon_c=\frac{T}{\nu}
\end{equation}
are overdamped.

Note that our analysis was assuming the so-called collisionless regime, where pumping the mode
 with a given energy $\epsilon_p$ one does not disturb the equilibrium distribution
function $N_k$ for thermal excitations involved in the damping process. This means that the
 relaxation of their distribution function occurs on a time scale
$\tau_R\gg\hbar/\epsilon_p$. For the discussed Beliaev damping, the thermal excitations that are
 involved in the damping of the mode with energy $\epsilon_p$ also have energies
$\sim\epsilon_p$. So, we have $\tau_R\sim \Gamma_p^{-1}$, and excitations with energies
 $\epsilon_p\gg\epsilon_c$ are well in the collisionless regime. However, for
$\epsilon_p\lesssim\epsilon_c$ we have the condition $\tau_R\epsilon_p/\hbar\lesssim 1$, and
 these excitations enter the hydrodynamic regime (see \cite{PS}). A dimensional
estimate for their damping rate is $\Gamma\sim \Gamma_p(\epsilon_p\tau_R/\hbar)$, with
 $\Gamma_p$ being the damping rate in the collisionless regime. Thus, for the modes with
energies $\epsilon_p\lesssim\epsilon_c$ we have the damping rate approaching $\epsilon_p$, i.e. they are significantly damped.

It should be noted that at finite temperatures we also have the Landau damping in which a given
 excitation with momentum ${\bf p}$ interacts with a thermal excitation (momentum
${\bf k}$), both are annihilated and an excitation with a higher energy and momentum
 (${\bf q}$) is created. There is a reversed process as well. The interaction Hamiltonian that
causes this damping is:
$$\hat V_L=g\sqrt{n}\sum_{{\bf k},{\bf q}}\int dxdy\left[f_0^*u_{{\bf q}}^*u_{{\bf k}}u_{{\bf p}}+2f_0^*\tilde v_{{\bf k}}^*\tilde v_{{\bf q}}u_{{\bf p}}-f_0\tilde v_{{\bf k}}^*
\tilde v_{{\bf p}}^*\tilde v_{{\bf q}}-2f_0u_{{\bf q}}^*\tilde v_{{\bf p}}^*u_{{\bf k}}\right]\,+h.c.,$$
and for the damping rate we have
$$\Gamma_{pL}=\frac{2\pi}{\hbar}\sum_{{\bf k},{\bf q}}|\langle {\bf q}|\hat V_L|{\bf p},{\bf k}\rangle|^2(N_k-N_q)\delta(\epsilon_q-\epsilon_p-\epsilon_k).$$
The calculations are similar to those made above for the Beliaev damping. In the low-energy limit
 in both limiting cases, $\epsilon_p\ll T$ and $\epsilon_p\gg T$, the results are
the same as in the case of Beliaev damping, but with a twice as small numerical coefficient.
 Thermal excitations that are involved in the damping of the mode with energy
$\epsilon_p$ also have energies $\sim\epsilon_p$. We thus see that the Landau damping does not
 change our conclusion made from the analysis of the Beliaev damping. Namely,
 excitations with energies $\epsilon_p\gg\epsilon_c$ are well in the collisionless regime, with the
  damping rate $\Gamma_p\ll\epsilon/\hbar$. On the other hand, excitations with
energies $\epsilon_p\lesssim\epsilon_c$ enter the hydrodynamic regime and are significantly damped.

For temperatures of the order of tens of nanokelvins and filling factors $\nu$ of the order of
 hundreds, like in experiments \cite{exp2,exp3} where the LLL regime has been reached,
Tkachenko modes with frequencies of the order of 1 Hz or lower should already be in the regime
 of strong damping. The strong damping of Tkachenko modes in this frequency range has
been observed in the JILA experiment \cite{exp1}, although this experiment was not yet in the LLL regime.

\section{One-body density matrix}

We now discuss correlation properties of rapidly rotating bosons in the mean-field Quantum Hall
 regime at zero and finite temperatures. For this purpose we will use the field
operators in the form (\ref{dfr}). Due to small fluctuations of the density the one-body density matrix takes the form:
\begin{equation}               \label{g1gen}
g_1({\bf r}) = \langle\hat\psi^{\dagger}({\bf r})\hat\psi(0)\rangle=\Psi_0^*({\bf r})\Psi_0(0)\exp\left\{-\frac{1}{2}\langle(\delta\hat\Phi({\bf
r})-\delta\hat\Phi(0))^2\rangle\right\}.
\end{equation}
In the low-momentum limit where $kl\ll 1$, omitting the term $ik_+l/2$  in the argument of $\Psi_0$ in equations (\ref{uB}) and (\ref{vB}),
the operator of the phase fluctuations given by Eq.~(\ref{df}) becomes:
\begin{equation}            \label{df1}
\delta\hat\Phi({\bf r})=-\frac{i}{2}\sum_{{\bf k}}\frac{(c_{1{\bf k}}+c_{2{\bf k}})}{\sqrt{N}}\exp(i{\bf k}{\bf r}/2)\,\hat a_{{\bf k}}\,+h.c.
\end{equation}
Then, using equations (\ref{c1k}) and (\ref{c2k}) for the mean square fluctuations we obtain:
\begin{equation}             \label{mspflgen}
\langle(\delta\hat\Phi({\bf r})-\delta\hat\Phi(0))^2\rangle=\alpha g\int\frac{d^2k}{(2\pi)^2}\,\frac{(1+2N_k)}{\epsilon_k}[1-J_0(kr/2)],
\end{equation}
where $J_0$ is the Bessel function.

At $T=0$ using the low-energy spectrum (\ref{lowkspectrum}) equation (\ref{mspflgen}) immediately gives:
\begin{equation}        \label{mspflinterm}
\langle(\delta\hat\Phi({\bf r})-\delta\hat\Phi(0))^2\rangle_0=\frac{2}{\sqrt{\eta}}\,\frac{1}{\nu}\int_0^{p_0}\frac{dk}{k}\,[1-J_0(kr/2)].
\end{equation}
The upper bound of the integration in Eq.~(\ref{mspfl0}) is $p_0\sim 1/l$, which represents the boundary of the first Brilluen zone, and for $r\gg l$ we find
\begin{equation}           \label{mspfl0}
\langle(\delta\hat\Phi({\bf r})-\delta\hat\Phi(0))^2\rangle_0\simeq \frac{2}{\sqrt{\eta}}\,\frac{1}{\nu}\ln\left(\frac{e^C r}{2l}\right),
\end{equation}
with $C=0.5772$ being the Euler constant.
For the  density matrix we then have an algebraic decay at large distances:
\begin{equation}            \label{g10}
g_1(r)\propto\left(\frac{l}{r}\right)^{1/\sqrt{\eta}\pi nl^2},\,\,\,\,\,\,\,\,r\gg l,
\end{equation}
which reproduces the result of Ref.~\cite{baym3}. So, in the thermodynamic limit there is no
 long-range order even at $T=0$, and we are dealing with a phase-fluctuating
Bose-condensed state.

At finite temperatures we have to take into account thermal fluctuations of the phase. Before
 finding $g_1({\bf r})$ we calculate the mean square fluctuations of the density
$\langle(\delta\hat n({\bf r})-\delta\hat n(0))^2\rangle/n^2$, which should be small for the validly
 of the mean-field approach. This is the case at $T=0$, but at
finite temperatures thermal fluctuations drastically change the situation. Using equations (\ref{dn}), (\ref{uB}), and (\ref{vB}) in the low-momentum limit we have:
\begin{equation}              \label{dnmeansquare1}
\frac{\langle(\delta\hat n({\bf r})-\delta\hat n(0))^2\rangle}{n^2}=\int_{0<k<l^{-1}}\frac{d^2k}{(2\pi)^2}\,\frac{g\alpha k^2l^2}{2\epsilon_k}[1-J_0(kr/2)](1+2N_k).
\end{equation}
For obtaining this relation we put $c_{1k}=c_{2k}=\alpha$ and expanded $\Psi_0(z+ik+l/2)$ in
 powers of $k$ up to the first order. Then, omitting small vacuum fluctuations and
writing $N_k\simeq T/\epsilon_k$ we reduce Eq.~(\ref{dnmeansquare1}) to
\begin{equation}        \label{dnmeansquare2}
\frac{\langle(\delta\hat n({\bf r})-\delta\hat n(0))^2\rangle}{n^2}=\frac{4T}{\alpha\eta ng\nu}\int_0^{\tilde l^{-1}}[1-J_0(kr/2)]\frac{dk}{k},
\end{equation}
where $\tilde l=l$ for $T\gg ng$, and for $T\ll ng$ we have $\tilde l=k_T^{-1}$ with the
 momentum $k_T$ following from the condition $\epsilon_{k_T}=T$.
The integration is straightforward and it yields:
\begin{equation}               \label{dnmeansquare}
\frac{\langle(\delta\hat n({\bf r})-\delta\hat n(0))^2\rangle}{n^2}\simeq\frac{4T}{\alpha\eta ng\nu}\ln\left(\frac{r}{2\tilde l}\right),
\end{equation}
i.e. the density fluctuations grow logarithmically with the distance. This means that at finite
 temperatures the mean-field approach can be employed only on a distance scale
$r<r_0=2\tilde l\exp(\alpha\eta ng\nu/4T)$. Then the mean square fluctuations of the density are
 small. In other words, the state is ordered in the vortex lattice only at $r<r_0$.

This introduces a low-momentum cut-off $r_0^{-1}$.  For the thermal mean square fluctuations of the phase  we then have:
\begin{equation}        \label{mspflintermT}
\langle(\delta\hat\Phi({\bf r})-\delta\hat\Phi(0))^2\rangle=
\frac{2\alpha g}{\pi}\int_{r_0^{-1}}^{l^{-1}}\frac{kdk}{\epsilon_k}\,\frac{[1-J_0(kr/2)]}{\exp (\epsilon_k/T)-1}.
\end{equation}
The most important contribution to the integral comes from low momenta, so that we can
 expand the exponent in the denominator of Eq.~(\ref{mspflintermT}) and put the upper limit of
integration equal to infinity.  Then Eq.~(\ref{mspflintermT}) takes the form:
\begin{equation}          \label{ffT}
\langle(\delta\hat\Phi({\bf r})-\delta\hat\Phi(0))^2\rangle= \frac{32 T }{\alpha \eta n g \nu}
\int_{r_0^{-1}}^{\infty}\frac{dk}{k^3}[1-J_0(kr/2)],
\end{equation}
A straightforward integration  at distances $r\gg l$ yields:
\beq
 \langle(\delta\hat\Phi({\bf r})-\delta\hat\Phi(0))^2\rangle= \frac{8T }{\alpha \eta n g \nu} \frac{r^2}{l^2} \ln\left(\frac{r_0}{r}\right).
\label{ff}
\eeq
Omitting vacuum phase fluctuations we then obtain an exponential decay of the density matrix:
\begin{equation}         \label{g1finalT}
g_1(r)\propto \exp\left[-\frac{4T}{\alpha\eta ng \nu}\frac{r^2}{l^2}\ln\left(\frac{r_0(T)}{r}\right)\right].
\end{equation}
For systems with a finite size $L<r_0$ we have to replace $r_0$ with $L$ in Eq.~(\ref{g1finalT}).

Stricktly speaking, equation (\ref{g1finalT}) is applicable only at distances $r\ll r_0$, where the
 system is ordered in the lattice. However, we clearly see that for $r$
approaching $r_0$
the density matrix practically drops to zero and, hence, it should remain close to zero at larger distances.

\section{Concluding remarks}


Concluding our work we would like to make a few remarks. First of all, the length scale $r_0$ on which the system is ordered in the vortex lattice at finite temperatures is
exponentially large and in cold atom experiments it exceeds the size of the sample. Even on approach to the melting point, assuming $T\simeq ng$ and $\nu\simeq 20$ we have
$r_0\simeq 300l$. However, the effect of finite temperature is likely to be dramatic for the visibility of the vortices. Equation (\ref{dnmeansquare}) shows that even fairly well
in the mean-field regime, for example at $\nu\simeq 40$ the mean square fluctuations of the density are $\langle(\delta\hat n({\bf r})-\delta\hat n(0))^2\rangle\sim 0.3$ for
$r\simeq 10l$ and $T\simeq ng$. This should significantly reduce the visibility of the vortices. We do not claim that these arguments explain the reduction of the vortex
visibility in the ENS experiment \cite{exp3}, but rather attract attention to this finite-temperature effect for future studies.

It is also worth mentioning that at finite temperatures the damping of Tkachenko modes may serve as a signature of the approach to the melting point of the lattice. The
characteristic excitation energy $\epsilon_c\simeq T/\nu$ below which these modes are strongly damped, increases significantly with decreasing the filling factor and becomes of the
order of $10$ Hz for $\nu=40$ even at temperatures as low as 20 nK.

\section*{Acknowledgements}

We are grateful to T. Jolicoeur, J. Dalibard, M.Yu. Kagan, and S. Ouvry for fruitful discussions and acknowledge support from the IFRAF Institute, from ANR (Grant 08-BLAN0165), and
from the Dutch Foundation FOM. This research has been supported in part by the National Science Foundation under Grant No. NSF PHYS05-51164. LPTMS is a mixed research unit No.
8626 of CNRS and Universit\'e Paris Sud.

\end{document}